\documentclass[sigconf]{acmart}
\renewcommand\footnotetextcopyrightpermission[1]{} 
\AtBeginDocument{%
  \providecommand\BibTeX{{%
    \normalfont B\kern-0.5em{\scshape i\kern-0.25em b}\kern-0.8em\TeX}}}

\makeatletter
\newcommand\footnoteref[1]{\protected@xdef\@thefnmark{\ref{#1}}\@footnotemark}
\makeatother

\usepackage{rotating,multirow}

\begin{document}

\title{``Subverting the Jewtocracy'': Online Antisemitism Detection Using Multimodal Deep Learning}


\author{Mohit Chandra}
\affiliation{%
  \institution{International Institute of Information Technology, Hyderabad}
  \city{Hyderabad}
  \country{India}}
\email{mohit.chandra@research.iiit.ac.in}

\author{Dheeraj	Pailla}
\authornote{Both authors contributed equally to this research.}
\affiliation{%
  \institution{International Institute of Information Technology, Hyderabad}
  \city{Hyderabad}
  \country{India}}
\email{dheerajreddy.p@students.iiit.ac.in}

\author{Himanshu Bhatia}
\authornotemark[1]
\affiliation{%
  \institution{International Institute of Information Technology, Hyderabad}
  \city{Hyderabad}
  \country{India}}
\email{himanshu.bhatia@students.iiit.ac.in}

\author{Aadilmehdi Sanchawala}
\affiliation{%
  \institution{International Institute of Information Technology, Hyderabad}
  \city{Hyderabad}
  \country{India}}
\email{aadilmehdi.s@students.iiit.ac.in}

\author{Manish Gupta}

\affiliation{%
  \institution{International Institute of Information Technology, Hyderabad}
  \city{Hyderabad}
  \country{India}}
\authornote{The author is also an applied researcher at Microsoft.}
\email{manish.gupta@iiit.ac.in}

\author{Manish	Shrivastava}
\affiliation{%
  \institution{International Institute of Information Technology, Hyderabad}
  \city{Hyderabad}
  \country{India}}
\email{m.shrivastava@iiit.ac.in}

\author{Ponnurangam Kumaraguru}
\affiliation{%
  \institution{IIIT, Delhi}
  \city{New Delhi}
  \country{India}}
\email{pk@iiitd.ac.in}
\renewcommand{\shortauthors}{Chandra, et al.}

\begin{abstract}
The exponential rise of online social media has enabled the creation, distribution, and consumption of information at an unprecedented rate. However, it has also led to the burgeoning of various forms of online abuse. Increasing cases of online antisemitism have become one of the major concerns because of its socio-political consequences.
Unlike other major forms of online abuse like racism, sexism, etc., online antisemitism has not been studied much from a machine learning perspective. To the best of our knowledge, we present the first work in the direction of automated multimodal detection of online antisemitism.
The task poses multiple challenges that include extracting signals across multiple modalities, contextual references, and handling multiple aspects of antisemitism.
Unfortunately, there does not exist any publicly available benchmark corpus for this critical task. Hence, we collect and label two datasets with 3,102 and 3,509 social media posts from Twitter and Gab respectively. 
Further, we present a multimodal deep learning system that detects the presence of antisemitic content and its specific antisemitism category using text and images from posts. 
We perform an extensive set of experiments on the two datasets to evaluate the efficacy of the proposed system. Finally, we also present a qualitative analysis of our study.
\end{abstract}




\maketitle

\section{Introduction}

Online social media (OSM) platforms have gained immense popularity in recent times due to their democratized nature, enabling users to express their views, beliefs, and opinions, and easily share those with millions of people. While these web communities have empowered people to express themselves, there have been growing concerns over the presence of abusive and objectionable content on these platforms. One of the major forms of online abuse which has seen a considerable rise on various online platforms is that of  antisemitism.\footnote{\url{http://www.crif.org/sites/default/fichiers/images/documents/antisemitismreport.pdf}} 

\begin{figure*}[!t]
\centering
\begin{minipage}{.48\textwidth}
  \centering
  \includegraphics[width=0.7\columnwidth]{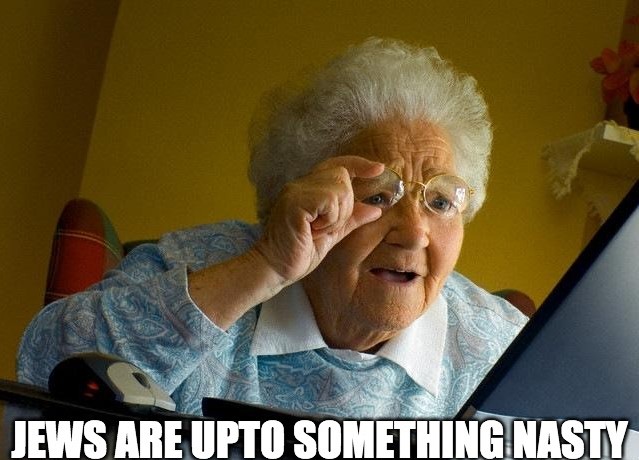}\\
  Text: Even grandma can see what's going on.
\end{minipage}%
\begin{minipage}{.48\textwidth}
  \centering
  \includegraphics[width=0.85\columnwidth]{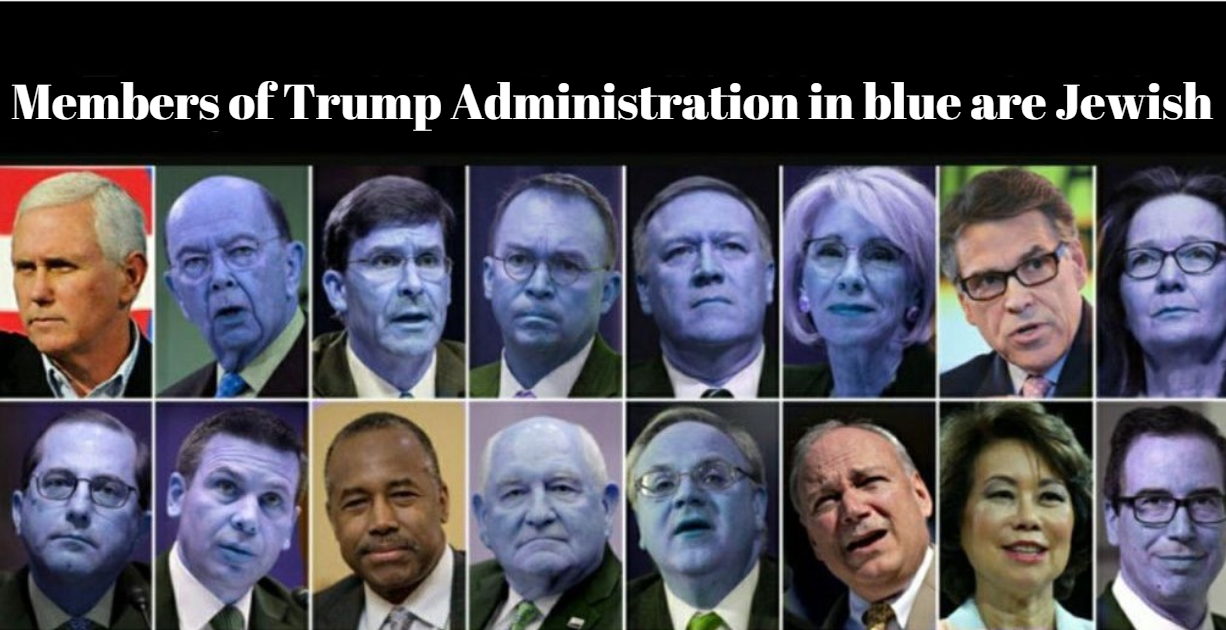}\\
  Text: I see the blews are at it again.
\end{minipage}
\caption{In both these examples, the post text appears to be non-antisemitic, due to absence of an explicit reference to Jews. But when looked along with the image, it can be classified as antisemitic.}
\label{fig:benign_example}
\end{figure*}

According to International Holocaust Remembrance
Alliance (IHRA)\footnote{\url{https://www.holocaustremembrance.com/}},  ``\textit{Antisemitism is a certain perception of Jews, which may be expressed as hatred toward Jews. Rhetorical and physical manifestations of antisemitism are directed toward Jewish or non-Jewish individuals and/or their property, toward Jewish community institutions and religious facilities.}''. Unlike some forms of hate-speech like sexism, cyberbullying, xenophobia etc., antisemitism originates from multiple aspects. In addition to the discrimination on the basis of race, antisemitism also includes discrimination on the basis of religion (e.g.difference from Christianity), economic activities (e.g.money lending) and political associations (e.g.Israel-Palestine issue, holding power at influential positions).

In recent times, online antisemitism has become one of the most widespread forms of hate-speech on major social media platforms\footnote{\url{https://ec.europa.eu/information_society/newsroom/image/document/2016-50/factsheet-code-conduct-8_40573.pdf}}. This recent trend of increased online hatred against Jews can also be correlated to the increasing real-world crime. According to the Anti-Defamation League's 2019 report, there has been a 12\% jump in the total cases of antisemitism amounting to a total of $2,107$ cases, and a disturbing rise of 56\% in antisemitic assaults as compared to 2018 across the U.S.\footnote{\url{https://www.adl.org/news/press-releases/antisemitic-incidents-hit-all-time-high-in-2019}}
Unlike studies on some major forms of online abuse like racism \cite{DBLP:conf/www/BadjatiyaG0V17}, cyber-bullying \cite{10.1145/3091478.3091487} and sexism \cite{parikh2019multi}, online antisemitism present on the web communities has not been studied in much detail from a machine learning perspective. This calls for studying online antisemitism in greater depth so as to protect the users from online/real world hate crimes.



In previous studies, it has been shown that real-world crimes can be related to incidents in online spaces~\cite{10.1007/s11280-017-0515-4}. Hence, it is essential to build robust systems to reduce the manifestation of heated online debates into real-world hate crimes against Jews. Due to the myriad of content on OSMs, it is nearly impossible to manually segregate the instances of antisemitism, thereby calling for the automation of this moderation process using machine learning techniques. Although antisemitic content detection is such a critical problem, there is no publicly available benchmark labeled dataset for this task. Hence, we gather multimodal data (text and images) from two popular social media platforms, Twitter and Gab. 

Oftentimes abusive content flagging policies across various social media platforms are very minimalistic and vague, especially for a specific abuse sub-category, antisemitism. Although various social media platforms have taken measures to curb different forms of hate-speech, more efforts are required to tackle the problem of online antisemitism.\footnote{\url{https://www.adl.org/holocaust-denial-report-card\#the-online-holocaust-denial-report-card-explained-}} As a result, we provide a detailed categorization methodology and label our datasets conforming to the same. Besides the dataset challenge, building a robust system is arduous because of the multimodal nature of the social media posts. A post with benign text may as well be antisemitic due to a hateful image (as shown in Fig.~\ref{fig:benign_example}). Thus, it becomes essential to take a more holistic approach rather than just inferring based on text. As a result, we take a multimodal approach which extracts information from text as well as images in this paper.

Fig.~\ref{fig:arch} shows a detailed architecture of our proposed multimodal antisemitism detection system which leverages the recent progress in deep learning architectures for text and vision. Given a (text, image) pair for a social media post, we use Transformer~\cite{vaswani2017attention}-based models to encode post text, and Convolutional Neural Networks (CNNs) to encode the image. Next, we experiment with multiple fusion mechanisms like concatenation, Gated MCB~\cite{fukui2016multimodal}, MFAS (Multimodal Fusion Architecture Search)~\cite{Perez-Rua_2019_CVPR} to combine text and image representations. The fused representation is further transformed using a joint encoder. The joint encoded representation is decoded to reconstruct the fused representation and also used to predict presence of antisemitism or an antisemitism category. 

We believe that the proposed work can benefit multiple stakeholders which includes -- 1) users of various web communities, 2) moderators/owners of social media platforms. Overall, in this paper, we make the following contributions: (1) We collect and label two datasets on online antisemitism gathered from Twitter and Gab with 3,102 and 3,509 posts respectively. Each post in both the datasets is labeled for presence/absence as well as antisemitism category. (2) We propose a novel multimodal system which learns a joint text+image representation and uses it for antisemitic content detection and categorization. The presented multimodal system achieves an accuracy of $\sim$91\% and $\sim$71\% for the binary antisemitic content detection task on Gab and Twitter respectively. Further, for 4-class antisemitism category classification, our approach scores an accuracy of $\sim$67\% and $\sim$68\% for the two datasets respectively, demonstrating its practical usability. (3) We provide a detailed qualitative study to analyse the limitations and challenges associated with this task and hate speech detection in general.


\section{Related Work}
\label{related works}
In this section, we present the past work, we broadly discuss-- (1) studies on antisemitism popular in sociology domain (2) popular hate speech datasets for Twitter and Gab, (3) deep learning studies for broad detection of hateful text content, and (4) overview of applications of multimodal deep learning.
\subsection{Previous Studies on Antisemitism}

Antisemitism as a social phenomenon has been extensively studied as part of social science literature~\cite{articleantisemitism,OAPENID:625675}. These studies have helped to explore the history behind antisemitism in-depth but lack quantitative analyses. Apart from the studies based on Sociology, there have been a few empirical studies. In one of the primary works, researchers collected around 7 million images and comments from `4chan' and `Gab' to study the escalation and spread of antisemitic memes in a longitudinal study~\cite{zannettou2020quantitative}.~\citet{doi:10.1177/2056305120916850} trained statistical machine learning models on data collected from twitter to detect antisemitic content. In contrast to this, our work focuses on detection of antisemitism through a multimodal deep learning framework. Recently, in an another work, researchers focused on detailed annotation analysis based on IHRA's guidelines for antisemitic content~\cite{jikeli2019annotating}. Our work on the other hand, encompasses the IHRA’s guidelines and incorporates fine-grained classes of antisemitism. Although antisemitism detection is such a critical problem, unfortunately, there hasn’t been any rigorous work on this problem from a deep learning perspective. We fill this gap in this paper.

\subsection{Hate Speech Datasets for Twitter and Gab}

Hatespeech detection has become a popular area for research and there have been quite a few works on dataset creation.~\citet{waseem2016hateful} annotated 16,914
tweets, including 3,383 as `sexist', 1,972 as `racist' and 11,559 as `neither'.~\citet{davidson2017automated} annotated $\sim$24K tweets for `hate speech', `offensive language but not hate' or `neither'. Another recent work presented a dataset comprising of $44,671$ posts from various social media platforms and annotated them as offensive or not. Gab was launched in 2016 and hence there are only a few dataset based studies.~\citet{qian-etal-2019-benchmark} proposed a hate-speech dataset on Gab with 33,776 posts annotated for hate versus non-hate.~\citet{chandra-etal-2020-abuseanalyzer} annotated 7,601 Gab posts for `Biased Attitude', `Act of Bias and Discrimination' or `Violence and Genocide'. Unlike the previous studies which have focused on general hate-speech or some popular sub-categories (like racism, sexism), we focus specifically on data related to antisemitism. 


\subsection{Deep Learning Methods for Detection of Various forms of Online Abuse} 

Deep learning has emerged as one of the most popular methods for hate-speech detection especially in Text-only NLP problems.~\citet{DBLP:conf/websci/FountaCKBVL19} proposed a Recurrent Neural Networks (RNN) based framework for classification of racism \& sexism, offensive speech, and cyberbullying using text and metadata. In contrast to this, we propose a general framework which doesn't require any metadata.~\citet{DBLP:conf/acl-alw/SerraLSSBV17} showed that character level based LSTMs (Long Short-Term Memory networks) can also be effective for abuse classification. In a more recent work,~\citet{parikh21_tweb} proposed a neural framework for classifying sexism and misogyny. 

Apart from LSTMs, Convolutional Neural Networks (CNNs) have been shown to be fairly successful for this task as they retain the spatial information to extract position invariant features.~\citet{gamback-sikdar-2017-using} used CNNs to classify the tweets into racist, sexist, both or none.~\citet{DBLP:conf/acl-alw/ParkF17} proposed a two-step hybrid approach for classification on hateful text into sexist or racist. They presented a hybrid CNN-based architecture which used sentence and word embeddings.~\citet{DBLP:conf/www/BadjatiyaG0V17} compared multiple deep learning architectures to classify tweets into racist, sexist, or neither. Unfortunately, there has been no previous work on exploring deep learning for antisemitism detection. Also, recently Transformer~\cite{vaswani2017attention} based methods have shown to outperform traditional deep learning methods like RNNs and LSTMs. Hence, we resort to methods like Bidirectional Encoder Representations from Transformers (BERT) and Robust BERT Approach (RoBERTa). Besides text, we also leverage semantics extracted from the accompanying image for improved antisemitism detection.

\subsection{Applications of Multimodal Deep Learning}
With the huge availability of multimodal data, multimodal deep learning has been harnessed to improve the accuracy for various tasks like Visual Question Answering (VQA)~\cite{Singh_2019_CVPR}, fake news/rumour detection~\cite{10.1145/3308558.3313552}, etc. Inspired by the success of Optical Character Recognition (OCR) on images for textVQA~\cite{Singh_2019_CVPR}, we also run OCR on the post images and use them for the classification task. Recently,~\cite{sabat2019hate,yang-etal-2019-exploring,Gomez_2020_WACV} have explored use of multimodal deep learning for general abuse detection from datasets like Reddit + Google images, and Twitter. Our proposed system differs from these previous methods in two important aspects: (1) they use traditional deep learning recurrent text methods like RNNs and LSTMs, while we investigate the application of more promising Transformer-based methods, (2) while previous methods were proposed for general hate, we focus on antisemitism.


\section{Antisemitism Categorization}
Besides annotating every post as antisemitic or not, we also annotate them for finer categories of online antisemitism. While there exists a good amount of literature exploring the ways in which antisemitism manifests itself, we primarily followed the categorization proposed by Brustein~\cite{brustein_2003} as it covers the major aspects of antisemitism. In his work, he explored the history behind the hatred against Jews and has categorized antisemitism into four categories, namely: (1) Political (2) Economic (3) Religious (4) Racial. We augmented this categorization with additional inputs from the detailed IHRA's\footnote{\url{https://www.holocaustremembrance.com/working-definition-antisemitism}} definition. We describe each category of antisemitism in detail in the following. 

\begin{table*}[!t]
    \centering
    \small
     \caption{Basic statistics for the two datasets.}
    \begin{tabular}{|l|c|c|c|c|c|c|}
    \hline
    &\#Total Posts&\#Antisemitic posts&\#Political posts&\#Economic posts&\#Religious posts&\#Racial  posts\\
         \hline
  Twitter&3,102&1,428&639&183&124&482\\
  \hline
Gab&3,509&1,877&736&118&144&879\\
    \hline
    \end{tabular}
   
    \label{tab:datasetStats}
\end{table*}

\begin{table*}[!t]
\small
\centering
\caption{Frequent Unigrams and Bigrams for each of the Antisemitism Categories.}
\begin{tabular}{|p{0.1\textwidth}|p{0.2\textwidth}|p{0.2\textwidth}|p{0.2\textwidth}|p{0.2\textwidth}|}
\hline
\textbf{N-Grams} & \textbf{Political Antisemitism} & \textbf{Economic Antisemitism} & \textbf{Religious Antisemitism} & \textbf{Racial Antisemitism}\\
\hline
\hline
Unigrams & jews, zionist, zog, israel, media, control. world, government, politics, conspiracy  & jewish, money, cash, finance, wealth, business, bankers, kosher & jews, christ, jesus, killer, rabbi, expel, satan, christians, messiah & jews, jewish, fake, holocaust, hitler, white, hebrew, ridiculous, pinocchio \\
\hline
Bigrams & world domination, zionist jews, zionist occupied, terrorist zionist, jews state & jewish money, money politics, money everything, money launderers, zionist bankers & christ killer, read torah, jesus killer, ultra orthodox, rabbi israel,  jewish ritual & jewish man, jews attacks, anti semitism, jewish people, race mixing\\
\hline
\end{tabular}
\label{tab:ngrams}
\end{table*}

\subsection{Political Antisemitism} 

Political Antisemitism can be defined as the hatred toward Jews based on the belief that Jews seek national and/or world power. In many of the cases lying in this category, Jews are portrayed to be controlling major political parties, governments, and decision making bodies. We also include the cases where they are accused of controlling media for promoting their interests (printing, Hollywood, etc). Furthermore, in some other cases, Jews are accused of being more loyal to Israel and blamed for the various socio-political crises. For example, \textit{The jews run congress through threats and intimidation.}

\subsection{Economic Antisemitism} 

Economic Antisemitism is based on the implicit belief that Jews perform and control the economic activities which are harmful for others or the society. This notion further exhibits multiple facets like Jews are undeservedly wealthy, greedy, dishonest, materialistic or cheaters. For example, \textit{the driving force behind globalism is jewish finance and greed.}

\subsection{Religious Antisemitism} 

Religious Antisemitism deals with bias and discrimination against Jews due to their religious belief in Judaism. Cases belonging to this category portray Jews as anti-Christ, Christ-killers, or against the teachings of the \textit{Bible}. Oftentimes, the posts also target Jewish religious institutions as well as their spiritual leader (\textit{Rabbi}). For example, \textit{may god strike down each and every one of these filthy jewish antichrists.}

\subsection{Racial Antisemitism} 

Unlike religious antisemitism, racial antisemitism is based on the prejudice against Jews as a race/ethnic group. Posts belonging to this category display a sense of inferiority for the Jewish race by portraying them as degenerate or attaching certain negative character as naturally inherited by them. Many posts in this category refer to false Jewish conspiracy for racial intermixing and blame them for LGBTQ+ related issues. Along with considering everything which discriminates Jews based on ethnic grounds in this category, we have also included instances talking about \textit{Holocaust} and its denial, since racial prejudice against Jews was one of the main cause for the aforementioned event.\footnote{\url{https://www.britannica.com/topic/anti-Semitism/Nazi-anti-Semitism-and-the-Holocaust}} For example, \textit{white is right which makes the jews always wrong}.

\begin{figure*}[!t]
\centering
  \includegraphics[width=0.9\textwidth]{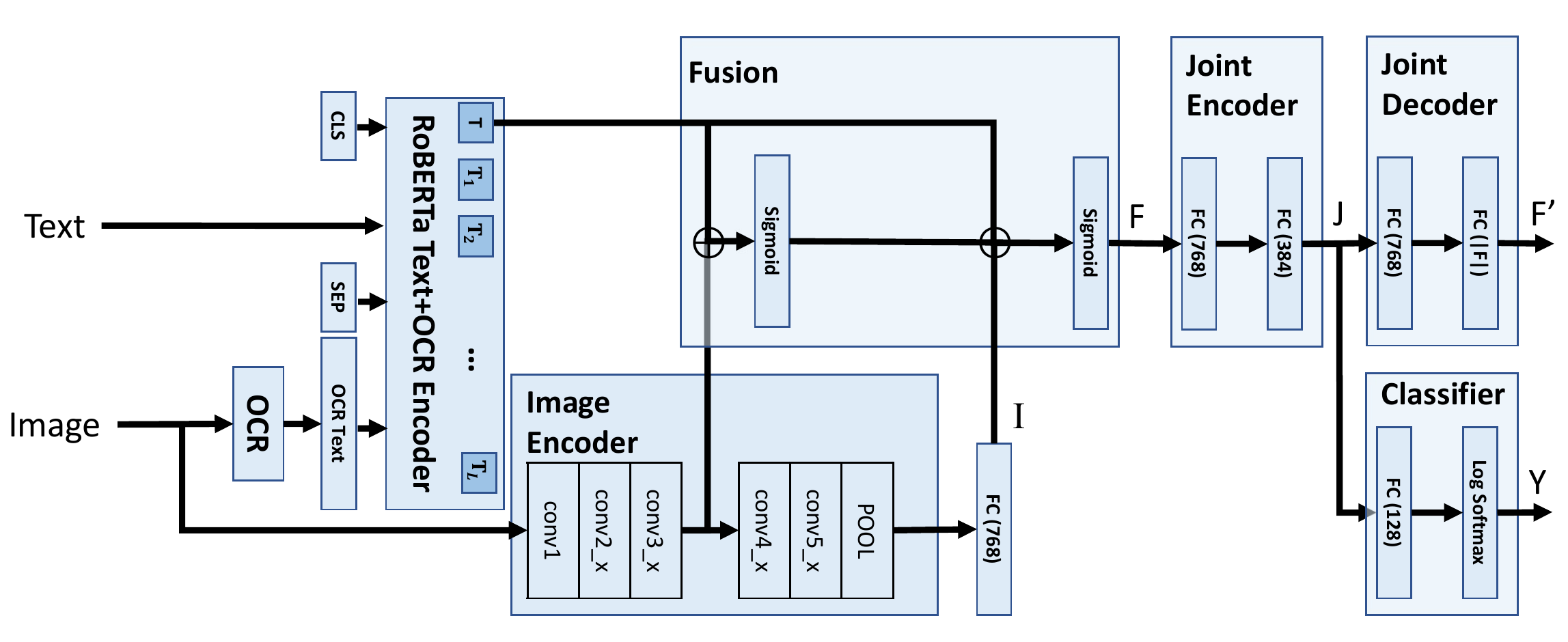}
  \caption{Proposed multimodal system architecture.}
  \label{fig:arch}
\end{figure*}

\section{Online Antisemitism Datasets}
We collect datasets from two popular social media platforms Twitter and Gab. In this section, we discuss details related to data collection, annotation and basic statistics.

\subsection{Data Collection}

We choose Twitter and Gab as the OSM platforms to gather data for our study. While Twitter has strict anti-abuse policies and an active content moderation team, Gab is an alt-right social media website with relaxed moderation policies. Recently, Gab has gained massive popularity especially among those who have been banned from mainstream web communities for violating their hate speech policy~\cite{DBLP:conf/www/ZannettouBCKSSB18}. After gathering a massive collection of posts from Twitter as well as Gab, we retained only those posts which contained text as well as images. Further, we ensured that each post included at least one term from a high precision lexicon.\footnoteref{note1} This lexicon contains common racial slurs used against Jews along with other words like `Jewish', `Hasidic', `Hebrew', `Semitic', `Judaistic', `israeli', `yahudi', `yehudi' to gather non-antisemitic posts as well, thereby helping maintain a balanced class distribution. Presence of these terms does not necessarily indicate presence/ absence of
antisemitism and hence we manually annotate the posts.

\subsection{Data Annotation}

For the annotation task, we selected four undergraduate students who are fluent in English. The annotators were given a detailed guideline along with examples to identify instances of antisemitism. Moreover, to ensure that the annotators had enough understanding of the task, we conducted multiple rounds of test annotations followed by discussions on disagreements. In the annotation procedure, each example was annotated by three annotators and the disagreements were resolved through discussion between all annotators. Each example was annotated on two levels after looking at the text as well as the image -- (1) binary label (whether the example in antisemitic or not), and (2) if the example is antisemitic then assign the respective antisemitism category. We used Fleiss’ Kappa score ~\cite{fleiss1971mns} to compute the inter-annotator agreement. The Fleiss' kappa score came out to be 0.707 which translates to a substantial agreement between the annotators. We removed all user sensitive information and followed other ethical practices to ensure user privacy.

\subsection{Data Statistics}

Table~\ref{tab:datasetStats} shows the post distribution across various classes for the two datasets. As observed, majority of posts in both datasets lie in either political antisemitism or the racial antisemitism category. We believe that this trend is inline with the phenomenon of `New antisemitism'.\footnote{\url{https://en.wikipedia.org/wiki/New_antisemitism}} Table~\ref{tab:ngrams} shows the frequent unigrams and bigrams for each of the antisemitism categories. Overall, 84\% of the total images had some form of text in them. This motivated us to use an OCR module in the proposed system. On average, post text has $\sim$45 and $\sim$27 words, while the OCR output is $\sim$50 and $\sim$51 words long, after pre-processing for Gab and Twitter respectively. We make the data publicly available.\footnote{\label{note1}\url{https://github.com/mohit3011/Online-Antisemitism-Detection-Using-MultimodalDeep-Learning}}

\section{Multimodal Antisemitism Categorization System}



Fig.~\ref{fig:arch} illustrates the architecture of our proposed multimodal system for online antisemitism detection with RoBERTa text+OCR encoder, ResNet-152 image encoder and the MFAS fusion module. Given a (text, image) pair for a social media post, we use Transformer~\cite{vaswani2017attention}-based models to encode (1) post text and (2) OCR text extracted from the image. We use CNNs to encode the image. Next, the fusion module combines the text and image representations. The fused representation is further transformed using a joint encoder. The joint encoded representation is decoded to reconstruct the fused representation and also used to predict presence of antisemitism or an antisemitism category. The entire network is trained end-to-end using back-propagation. Since our datasets are relatively small, we fine-tune the pre-trained networks on the presented datasets for multimodal classification. The code for the proposed system can be found here.\footnoteref{note1} We now describe each module in detail.

\begin{table*}[!t]
\small
\centering
\caption{Comparison of (5-fold cross validation) performance of popular text-only and image-only classifiers. The best performing method is highlighted in bold separately for both the text and image blocks.}
\begin{tabular}{|l|l|c|c|c|c|c|c|c|c|}
\hline
\multicolumn{2}{|c|}{}& \multicolumn{4}{|c|}{Twitter} & \multicolumn{4}{|c|}{Gab}\\
\cline{3-10}
\multicolumn{2}{|c|}{}& \multicolumn{2}{|c|}{Binary} & \multicolumn{2}{|c|}{Multiclass} &\multicolumn{2}{|c|}{Binary} & \multicolumn{2}{|c|}{Multiclass}\\
\cline{3-10}
\multicolumn{2}{|c|}{}& Accuracy & F-1 & Accuracy & F-1& Accuracy & F-1& Accuracy & F-1\\
\hline
\hline
\multirow{5}{*}{\rotatebox{90}{Text only}}&GloVe+Dense &.630$\pm$.009&.621$\pm$.013&.490$\pm$.013&.268$\pm$.025&.651$\pm$.027&.612$\pm$.040&.540$\pm$.031&.276$\pm$.018\\
\cline{2-10}
&FastText+Dense &.540$\pm$.000&.351$\pm$.000&.467$\pm$.031&.223$\pm$.099&.566$\pm$.017&.429$\pm$.045&.532$\pm$.030&.269$\pm$.019\\
\cline{2-10}
&GloVe+att-RNN&.583$\pm$.048&.552$\pm$.081&.416$\pm$.019&.239$\pm$.033&.630$\pm$.039&.624$\pm$.045&.460$\pm$.039&.240$\pm$.028\\
\cline{2-10}
&BERT+Dense &.701$\pm$.015&.700$\pm$.016&\textbf{.669}$\pm$\textbf{.047}&\textbf{.676}$\pm$\textbf{.036}&\textbf{.889}$\pm$\textbf{.008}&\textbf{.889}$\pm$\textbf{.009}&.623$\pm$.025&.575$\pm$.038\\
\cline{2-10}
&RoBERTa+Dense&\textbf{.733}$\pm$\textbf{.007}&\textbf{.733}$\pm$\textbf{.008}&.663$\pm$.039&.662$\pm$.050&.874$\pm$.010&.874$\pm$.010&\textbf{.632}$\pm$\textbf{.032}&\textbf{.583}$\pm$\textbf{.039}\\
\hline
\hline
\multirow{2}{*}{\rotatebox{90}{\vtop{\hbox{\strut Img}\hbox{\strut only}}}}&ResNet-152 &\textbf{.579}$\pm$\textbf{.014}&\textbf{.578}$\pm$\textbf{.015}&\textbf{.416}$\pm$\textbf{.028}&\textbf{.317}$\pm$\textbf{.040}&.587$\pm$.008&.583$\pm$.008&\textbf{.456}$\pm$\textbf{.020}&\textbf{.275}$\pm$\textbf{.010}\\
\cline{2-10}
&Densenet-161 &.567$\pm$.033&.566$\pm$.033&.405$\pm$.033&.281$\pm$.011&\textbf{.610}$\pm$\textbf{.017}&\textbf{.607}$\pm$\textbf{.015}&.446$\pm$.031&.274$\pm$.027\\
\hline
\end{tabular}
\label{tab:singleModalResults}
\end{table*}

\subsection{Text + OCR Encoder}

We remove URLs and non alpha-numeric characters. The cleaned text is tokenized and then encoded using the BERT/RoBERTa tokenizer and encoder respectively~\cite{Wolf2019HuggingFacesTS}. For getting the OCR output from the images we experimented with three different services (Google's Vision API, Microsoft's Computer Vision API and Open source tesseract). We found the Google's Vision API to perform the best for the broad range of images we had in the dataset (from newspaper articles to memes). The extracted OCR text goes through the same pre-processing process as the post text. We experiment with BERT and RoBERTa for text encoding since they have been shown to lead to high accuracy across multiple NLP tasks.

BERT~\cite{devlin2018bert} is a transformer encoder with 12 layers, 12 attention heads and 768 dimensions. We used the pre-trained model which has been trained on Books Corpus and Wikipedia using the MLM (masked language model) and the next sentence prediction (NSP) loss functions. The input to the BERT model is obtained as a concatenation as follows: (CLS, post text, SEP, OCR text) where CLS and SEP are the standard classification and separator tokens respectively. The 768-dimensional representation $T$ for the ``CLS'' token from the last encoder layer is used as input by the fusion module. 

RoBERTa~\cite{liu2019roberta} is a robustly optimized method for pretraining natural language processing
(NLP) systems that improves on BERT. RoBERTa was trained with much more data -- 160GB of text
instead of the 16GB dataset originally used to train BERT. It is also trained for larger number of iterations
up to 500K. Further, it uses larger byte-pair encoding (BPE) vocabulary with 50K subword units instead
of character-level BPE vocabulary of size 30K used for BERT. Finally, compared to BERT, it removes
the next sequence prediction objective from the training procedure, and a dynamically changing masking
pattern is applied to the training data.

\subsection{Image Encoder}

We perform Gaussian normalization for each image, and resize images to 224x224x3 size and feed them to a CNN. We augment our training data using image transformations such as random cropping and random horizontal flipping. We connect the output of second last layer from these networks to a 768 sized dense layer. Output from this layer $I$ is used as the image representation. 

Few CNN architectures are popular like: AlexNet~\cite{krizhevsky2012imagenet}, InceptionV3~\cite{szegedy2016rethinking}, VGGNet-19~\cite{simonyan2014very}, Resnet-152~\cite{7780459} and Densenet-161~\cite{Huang_2017}. We chose Resnet-152 and Densenet-161 since they have been shown to outperform the other CNN models across multiple vision tasks.

\subsection{Fusion}

To combine the features obtained from the Text + OCR and the image encoder modules, $T$ and $I$, we experiment with three different techniques of fusion -- (1) Concatenation (2) Gated MCB~\cite{fukui2016multimodal} and (3) MFAS~\cite{Perez-Rua_2019_CVPR}. Gated MCB (Multimodal Compact Bilinear) pooling combines multimodal features using an outer product followed by a sigmoid non-linearity . We also experimented with Hadamard inner product but found it to be worse compared to gated MCB, in line with previous literature on multimodal deep learning. As shown in Fig. 3 in~\cite{Perez-Rua_2019_CVPR}, MFAS (Multimodal Fusion Architecture Search) first concatenates text and image representations from an intermediate hidden layer, applies a sigmoid non-linearity, and then concatenates this with final layer text and image representations along with a sigmoid non-linearity.

\subsection{Joint Encoder/Decoder and Classifier}

The fused representation  $F$ is then passed through a series of Dense layers (768 and 384) to obtain a joint encoded vector $J$. $J$ is fed to two modules: joint decoder and classifier. The joint decoder again consists of dense layers of sizes 768 and $|F|$. The classifier feeds the output $J$ to a dense layer of size 128 and then finally to the output log-softmax layer. The joint decoder aims to reconstruct $F$ and uses MSE (mean squared error) loss, while the classifier aims to predict presence/absence of antisemitism or antisemitism category. We use the sum of these two losses to train the model.

\section{Experiments}

In this section, we discuss hyper-parameter settings; results using text-only, image-only and multimodal classifiers; and qualitative analysis using attention visualization, error analysis and case studies.

\subsection{Hyper-Parameter Settings for Reproducibility}

We use the following experimental settings. We perform 5 fold cross validation where we split our labeled data in 64:16:20 as our train, validation, test split for each fold. All hyper-parameters were tuned using validation set. For the MFAS fusion module, we use the block 2 output as the intermediate layer output since it gave us the best results compared to output from other blocks (on validation set). For Gated-MCB all experimental settings were used as suggested by the reference paper. For MFAS, $|F|=|F'|$=2,816 (which is 3*768+512); for other fusion methods $|F|=|F'|$=1,536 (which is 2*768).

For all experiments, we use Adam optimizer~\cite{kingma2014adam}. We experimented with a range of learning rates and found $lr = 2e^{-6}$ as the best one. To improve the stability of the system across the samples we used a batch normalization layer before the Dense layers. For the Dense layers, we use dropouts with a drop probability of $0.2$. We used RELU non-linearity after all our dense layers except the final output layer. We train our system for a max of 50 epochs with early stopping, with a batch size of 4. For all the results, we report 5-fold cross-validation accuracy and macro-F1. For further details of hyper-parameters, we refer the reader to look at our codebase\footnote{\url{https://github.com/mohit3011/Online-Antisemitism-Detection-Using-MultimodalDeep-Learning}}. 


\begin{table*}[!t]
\small
\centering
\caption{Comparison of (5-fold cross validation) performance of multimodal classifiers with RoBERTa as text encoder and ResNet-152 as image encoder. We also compare the performance of the proposed architecture with a baseline from Gomez et al.\cite{Gomez_2020_WACV} (FCM).}
\begin{tabular}{|l|c|c|c|c|c|c|c|c|}
\hline
& \multicolumn{4}{|c|}{Twitter} & \multicolumn{4}{|c|}{Gab}\\
\cline{2-9}
Method&\multicolumn{2}{|c|}{Binary} & \multicolumn{2}{|c|}{Multiclass} &\multicolumn{2}{|c|}{Binary} & \multicolumn{2}{|c|}{Multiclass}\\
\cline{2-9}
& Accuracy & F-1 & Accuracy & F-1& Accuracy & F-1& Accuracy & F-1\\
\hline
\hline
FCM~\cite{Gomez_2020_WACV}&.564$\pm$.015&.545$\pm$.038&.445$\pm$.006&.164$\pm$.022&0.607$\pm$0.014&.595$\pm$.028&.468$\pm$.005&.182$\pm$.027\\
\hline
Concatenation&.710$\pm$.012&.708$\pm$.013&.662$\pm$.027&.664$\pm$.027&.905$\pm$.005&.905$\pm$.005&.653$\pm$.052&.616$\pm$.046\\
\hline
Gated MCB&.690$\pm$.026&.683$\pm$.036&.679$\pm$.030&\textbf{.677}$\pm$\textbf{.043}&.904$\pm$.014&.903$\pm$.014&.654$\pm$.039&.618$\pm$.043\\
\hline
MFAS&\textbf{.715}$\pm$\textbf{.013}&\textbf{.714}$\pm$\textbf{.014}&\textbf{.680}$\pm$\textbf{.035}&.675$\pm$.023&\textbf{.906}$\pm$\textbf{.007}&\textbf{.906}$\pm$\textbf{.007}&\textbf{.665}$\pm$\textbf{.029}&\textbf{.625}$\pm$\textbf{.032}\\
\hline
\end{tabular}
\label{tab:multiModalResults}
\end{table*}

\begin{figure*}[!t]
\centering
\begin{minipage}{.65\textwidth}
  \centering
  \includegraphics[width=\linewidth]{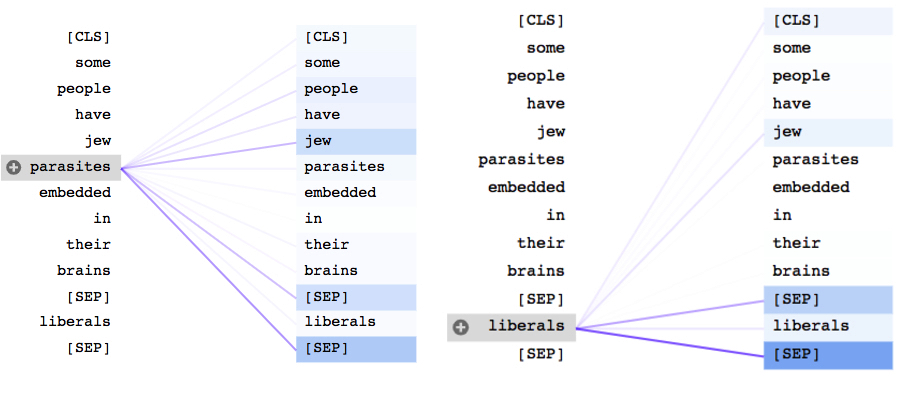}
  \captionof{figure}{Text + OCR encoder module attention visualization}
  \label{fig:bertviz_visual}
\end{minipage}%
\begin{minipage}{.3\textwidth}
  \centering
  \includegraphics[width=0.9\linewidth]{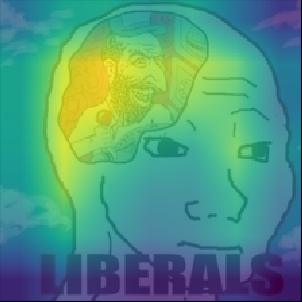}
  \captionof{figure}{Image encoder module attention visualization (Best viewed in color)}
  \label{fig:cmap_visual}
\end{minipage}
\end{figure*}

\subsection{Results using Text-only and Image-only Classifiers}

We experiment with five popular pre-trained text embedding/network based classifiers and two pre-trained image network classifiers. For the text-only classifiers we use GloVe~\cite{DBLP:conf/emnlp/PenningtonSM14}, FastText~\cite{joulin2016fasttext}, BERT~\cite{devlin2018bert} and RoBERTa~\cite{liu2019roberta}. We also experiment with Founta et al.~\cite{DBLP:conf/websci/FountaCKBVL19}'s method which is an attentional RNN model with GloVe embeddings. For the image only classifiers we experiment with ResNet-152~\cite{7780459} and DenseNet-161~\cite{Huang_2017}. We also experimented with VGG-19 but did not see any better results.

Table~\ref{tab:singleModalResults} provides the comparative results. We make the following observations:
(1) Compared to the text-only methods, the image-only models provide much lower accuracy. We believe this is because images related to antisemitic posts are usually memes, screenshots, or news articles that usually don't carry any spatial-visual features. (2) Among the text-only classifiers, Transformer based methods performed better than the rather shallow approaches like FastText and GloVe. BERT and RoBERTa lead to very similar results. (3) Among the image-only models, ResNet-152 performs the best except for binary classification on Gab. 

\subsection{Results using Multimodal Classifiers}

In this experiment we tested different fusion mechanism for our proposed multimodal classifier. From Table~\ref{tab:singleModalResults}, we observe that ResNet-152 is the best image encoder and RoBERTa is the best text (post text + OCR) encoder. Hence, we perform multimodal experiments with these encoders only. We show the results obtained for this multimodal experiment in Table~\ref{tab:multiModalResults}. In addition to this, we also compared the performance of our proposed architecture with the baseline model from~\cite{Gomez_2020_WACV} (FCM). FCM uses GloVe for text encoder and InceptionV3 for image encoder.

We make the following observations: (1) Each of the three variants of the proposed architecture beat the baseline by a huge margin. (2) Results in Table~\ref{tab:multiModalResults} are much better compared to those in Table~\ref{tab:singleModalResults} except for the binary classification for Twitter. For Gab, we see a massive increase of $\sim$2 and $\sim$4 percentage points in accuracy and F1 for both binary and multi-class tasks  respectively. The improvement in results when using both text and images (i.e., across Tables~\ref{tab:multiModalResults} and~\ref{tab:singleModalResults}) is better for Gab overall compared to Twitter. This is because the images on Gab are much more rich and informative compared to those on Twitter. (3) MFAS based multimodal fusion approach outperforms the Gated MCB and concatenation based fusion approaches.


\subsection{Qualitative analysis: Attention Visualization}

To gain better insights into the the proposed system, we visualize  attention weights for both the Text + OCR (using bertviz~\cite{vig2019multiscale}) and the Image encoder (using GradCAM~\cite{selvaraju2017grad}). Figure \ref{fig:bertviz_visual} shows the visualization for a self-attention head from the last encoder layer in the Text + OCR module. We took an antisemitic example having the text content as ``\textit{some people have jew parasites embedded in their brains}'' and the OCR text being ``\textit{liberals}". We observe  high attention between the word `parasites' with `jew' apart from the standard RoBERTa [SEP] tokens showcasing that the system identified that this text refers Jews as parasites. Similarly, in Fig.~\ref{fig:bertviz_visual} (right), the word `liberals' present in the OCR text output shares higher attention weights with the word `jew' from the post text content showing the cross-attention learnt by the system. 

Fig.~\ref{fig:cmap_visual} shows the GradCAM visualization of the image in the post, we observe higher attention on the region of the \textit{Happy Merchant Meme} face which is usually used as a symbol of antisemitism. Another interesting observation is that the text in the image doesn't get much attention, which makes our choice of adding OCR suitable.

\begin{table*}[!t]
\centering
    \small
    \caption{Top: Correctly predicted examples. Bottom: Examples with erroneous predictions.}
    \begin{tabular}{|p{0.25\textwidth}|p{0.25\textwidth}|p{0.1\textwidth}|p{0.1\textwidth}|p{0.2\textwidth}|}
\hline
Post text&OCR Text/Image Description&Actual Class&Predicted Class&Explanation\\
\hline
\hline
shabbat shalom to all my jewish friends may the lord bless you&shabbat shalom everyone &Non-Antisemitic&Non-Antisemitic&The terms ``friends'', ``Shabbat'', ``Shalom'' are good clues.\\
\hline
no more jewish wars for israel& I see dead people wherever jews have the power&Antisemitic&Antisemitic&The terms ``dead'', ``jewish'' and ``wars'' are good clues.\\
\hline
\hline
Zog (2020): The heartwarming story of a magical dragon who eventually takes control of the entertainment industry.& ZOG (with a picture of a dragon)&Antisemitic&Non-Antisemitic&This post presents a case of sarcasm where ZOG (the dragon cartoon) is used to refer zionist occupied government (ZOG)\\
\hline
Beautiful woman. Not this are zionist woman. They have weapons everywhere. & (No Text)&Racial Antisemitism&Political Antisemitism& The presence of word `zionist' causes confusion\\
\hline
Banksters jews and the blood from white people&(image with people carrying money bags and dead people)&Economic Antisemitism&Racial Antisemitism&Reference to `white people' causes confusion.\\
\hline
    \end{tabular}
    \label{tab:examples}
\end{table*}

\subsection{Qualitative analysis: Error Analysis and Case Studies}

\begin{table}[ht]
    \small
    \centering
    \caption{Confusion matrix for the binary classification task (Gab). The entries represent the sum on test set examples over 5-fold cross validation.}
    \begin{tabular}{|l|l|l|l|}
    \hline
    \multirow{5}{*}{\rotatebox{90}{\small Actual}}&&\multicolumn{2}{c|}{Predicted}\\
    \cline{3-4}
    &&Non-antisemitic&Antisemitic\\
    \cline{3-4}
    \cline{3-4}
    &Non-antisemitic&1470&162\\
    \cline{2-4}
    &Antisemitic&167&1710\\
    \hline
    \end{tabular}
    
    \label{tab:cm_binaryGab}
\end{table}

\begin{table}[ht]
    \centering
    \caption{Confusion matrix for the multiclass classification task (Gab). The entries represent the sum on test set examples over 5-fold cross validation.}
    \begin{tabular}{|l|l|l|l|l|l|}
    \hline
    \multirow{6}{*}{\rotatebox{90}{\small Actual}}&&\multicolumn{4}{c|}{Predicted}\\
    \cline{2-6}
    &&Political&Economic&Religious&Racial\\
    \cline{2-6}
    \cline{2-6}
    &Political&441&49&33&213\\
    \cline{2-6}
    &Economic&14&82&3&19\\
    \cline{2-6}
    &Religious&9&1&102&32\\
    \cline{2-6}
    &Racial&141&40&76&622\\
    \cline{2-6}
    \hline
    \end{tabular}
    
    \label{tab:cm_multiGab}
\end{table}

Tables~\ref{tab:cm_binaryGab} and~\ref{tab:cm_multiGab} show the confusion matrices for the proposed MFAS-based multimodal system for binary and multi-class cases respectively for Gab. Similarly,  Tables~\ref{tab:cm_binaryTwitter} and~\ref{tab:cm_multiTwitter} show confusion matrices for Twitter. Each entry in the confusion matrices represents the sum of examples in the test sets over 5-fold cross validation. As observed in Table~\ref{tab:cm_binaryGab} and Table~\ref{tab:cm_binaryTwitter}, the classifier has higher percentage of \textit{False Positives} than the \textit{False Negatives} (where the positive class is `Antisemitic'). We believe that this was due to many borderline cases which confused the classifier on topics like `\textit{Anti-Israel hate}', `\textit{Issue of Israel-Palestine conflict}' etc. Additionally, from Table~\ref{tab:cm_multiGab} and Table~\ref{tab:cm_multiTwitter} we observe that across both the datasets, the classifier is most confused between the `Political' and `Racial' classes. This could be because many politically oriented posts against Jews also used racial prejudices.  

\begin{table}[ht]
    \centering
    \caption{Confusion matrix for the binary classification task (Twitter). The entries represent the sum on test set examples over 5-fold cross validation.}
    \begin{tabular}{|l|l|l|l|}
    \hline
    \multirow{5}{*}{\rotatebox{90}{\small Actual}}&&\multicolumn{2}{c|}{Predicted}\\
    \cline{3-4}
    &&Non-antisemitic&Antisemitic\\
    \cline{3-4}
    \cline{3-4}
    &Non-antisemitic&1106&568\\
    \cline{2-4}
    &Antisemitic&317&1111\\
    \hline
    \end{tabular}
    
    \label{tab:cm_binaryTwitter}
\end{table}

\begin{table}[ht]
    \centering
    \caption{Confusion matrix for the multiclass classification task (Twitter). The entries represent the sum on test set examples over 5-fold cross validation.}
    \begin{tabular}{|l|l|l|l|l|l|}
    \hline
    \multirow{6}{*}{\rotatebox{90}{\small Actual}}&&\multicolumn{4}{c|}{Predicted}\\
    \cline{2-6}
    &&Political&Economic&Religious&Racial\\
    \cline{2-6}
    \cline{2-6}
    &Political&470&35&11&123\\
    \cline{2-6}
    &Economic&16&149&9&9\\
    \cline{2-6}
    &Religious&15&4&79&26\\
    \cline{2-6}
    &Racial&160&12&37&273\\
    \cline{2-6}
    \hline
    \end{tabular}
    
    \label{tab:cm_multiTwitter}
\end{table}

Finally, in Table~\ref{tab:examples}, we present a few examples where our system produced correct/incorrect (top/bottom part) predictions. The last ``Explanation'' column details the plausible reason for the erroneous cases.

\begin{figure}[ht]
    \centering
        \includegraphics[width=\columnwidth]{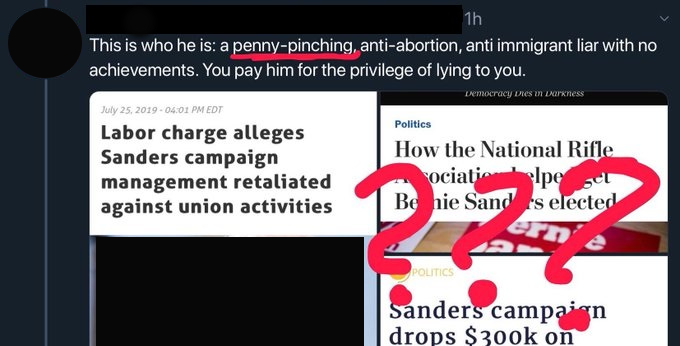}
      \caption{In this example, the image contains screenshot of multiple tweets/posts/articles stitched together posted by someone else.}
      \label{fig:image_error_1}
\end{figure}
\begin{figure}[ht]
    \centering
    \includegraphics[height=2in]{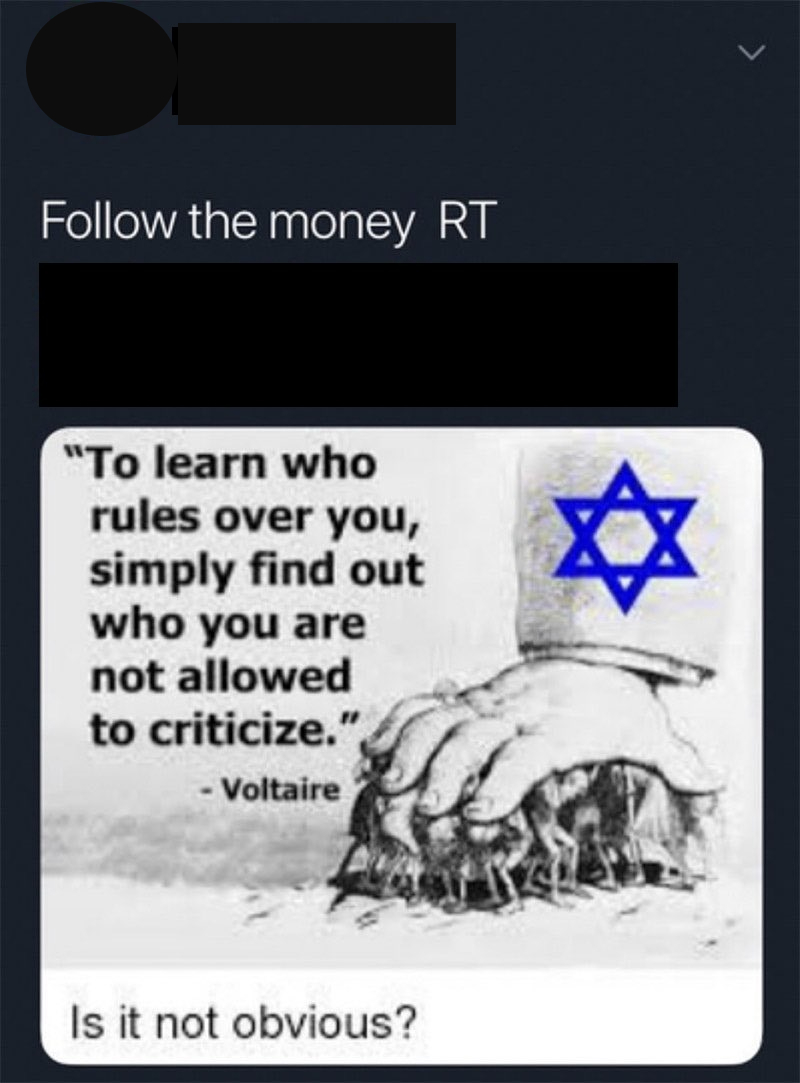}
      \caption{In this example, the image is a screenshot of a hateful tweet posted by someone else.}
      \label{fig:image_error_2}
\end{figure}

Figures~\ref{fig:image_error_1} and~\ref{fig:image_error_2} present two interesting instances where our multimodal system misclassifies. The post referred in Figure~\ref{fig:image_error_1} had the post text as ``\textit{calling a Jewish man a penny pincher is anti semitic}''. The post is condemning antisemitic behaviour but the complex structure of text present in the image makes it hard for the model to extract information correctly. Also, the model does not understand that the original tweet was posted by some other user and not this user. Similarly, the post referred in Figure~\ref{fig:image_error_2} had the post text as ``\textit{@usermention teams up with another antisemite this time a guy who tweeted an image depicting Jews as controlling the world and adding that Jewish money crushes the little people}''. This post reports an antisemitic behaviour by someone else through the screenshot of the tweet. But, due to the absence of any additional context about the image being a screenshot of the tweet from someone else, makes the system to commit the mistake. These two cases help us surface a broader problem with the current systems capturing information from multiple modalities. 

\section{Limitation}

The presented task of antisemitism detection share similar set of limitations as with other hate speech detection tasks like racism/sexism detection. We list a few of the limitations here:
\begin{itemize}
    \item \textbf{Keyword Bias:} As with the other hate speech detection systems, keywords play an important role in the classification of content. We observed that posts containing  certain keywords like \textit{zionists, holocaust, Hitler, Christians, Torah} were prone to be classified as antisemitic since majority of training posts containing these keywords were labelled as antisemitic. This observation is in line with the past research which claims that deep learning models learn this kind of keyword biases.
    \item \textbf{Subtlety in the expression of hate:} Another set of examples which were misclassified belonged to the category of sarcasm/trolling/subtle hate. It becomes extremely difficult for the system to extract the real intent behind posts expressing views in a subtle manner. This in turn creates a dilemma of freedom of speech/curbing hate speech which is a common problem across various other forms of hate speech.
    \item \textbf{Noise from multiple modalities}: Though we showed that adding information from multiple modalities overall helps in antisemitism detection, in a few cases noise present in one of the modalities caused misclassification (as shown in Figures~\ref{fig:image_error_1} and~\ref{fig:image_error_2}).
\end{itemize}

\section{Discussion}
In this work we presented the first systematic study on the problem of detection and categorization of antisemitism. We collected and labeled two datasets for antisemitism detection and categorization. We hope that these will accelerate further research in this direction. We proposed a multimodal system which uses text, images and OCR for this task and demonstrated its efficacy on the two datasets. We experimented with single-modal as well as multimodal classifiers and found that combining data from multiple modalities improves the performance and robustness of the system to a small extent for Twitter but massively for Gab. Finally, we also performed a qualitative analysis of our multimodal system through attention visualisation and error analysis. We observed that the complexities in images and subtlety of hate in text can lead to errors. Images with multiple screenshots, multi-column text and texts expressing irony, sarcasm or indirect references posed problems for the classifier.

Similar to images, videos have become increasingly common. It will be interesting to develop multimodal systems involving text and videos for detecting antisemitism in the future. Another interesting direction involves usage of contextual information like user profiles for the classification task.

\bibliographystyle{ACM-Reference-Format}
\bibliography{references}
\end{document}